\documentclass{aastex631}

\usepackage{bm}
\usepackage{subcaption}
\usepackage[T1]{fontenc}
\usepackage{booktabs}
\usepackage{threeparttable}
\usepackage{multirow}
\usepackage{amsmath}
\usepackage[utf8]{inputenc}
\usepackage{makecell}
\usepackage{hyperref}
\usepackage{numprint}
\usepackage[ruled,vlined]{algorithm2e}
\usepackage{tabularx}

\begin{document}

 \title{Exploring Braess' paradox in pedestrian evacuation traffic: experiment and modeling}
 
\author{Jinghui Wang}

\affiliation{School of Safety Science and Emergency Management\\
Wuhan University of Technology\\
Wuhan, China}

\author{Wei Lv}
\altaffiliation{{\url{weil@whut.edu.cn} (W. Lv)}}
\affiliation{School of Safety Science and Emergency Management\\
Wuhan University of Technology\\
Wuhan, China}
\affiliation{China Research Center for Emergency Management\\
Wuhan University of Technology\\
Wuhan, China}

\begin{abstract}
The emergence of the Braess' paradox in road traffic systems demonstrates the positive effect of transportation planning in improving efficiency. By contrast, the phenomenon has rarely been examined in pedestrian evacuation traffic. Yet the possibility that Braess’ paradox could lengthen evacuation times under hazardous conditions has received little systematic attention. In this paper, we investigate Braess' paradox in pedestrian evacuation traffic through a series of supervised experiments and corresponding traffic assignment models to examine its potential occurrence. Our empirical and modeling results indicate that Braess’ paradox is unlikely to be a prevalent phenomenon in pedestrian traffic systems. Specifically, under autonomous evacuation and the assumption of complete network knowledge, the paradox does not arise in high-demand evacuation contexts in our case studies. Under a more realistic assumption of limited network knowledge, however, the paradox can occur. These findings highlight the importance of information conditions for evacuation performance and provide guidance for the design and management of large public venues.

\end{abstract}


\section{Introduction} 
\label{section1}

The Crystal Palace erected for the Great Exhibition of 1851 is widely regarded as the point of departure for modern large public buildings \citep{banham2020megastructure}. Driven by cultural, transport, and leisure needs, large public facilities such as airports, railway stations, stadiums, shopping malls, and theaters have become integral to the functioning of the city. In anticipation of substantial demand, these venues are frequently conceived as very large integrated complexes that accommodate multiple functions, with numerous entrances and exits and dense networks of interconnected corridors designed to support efficient pedestrian movement. The emergence of such large public buildings has required parallel advances in planning, design, and management. A concomitant concern is pedestrian safety, and lots cases have demonstrated that inappropriate design and management failures can lead to catastrophic results \footnote{\url{https://en.wikipedia.org/wiki/1994_Karamay_fire}}\footnote{\url{https://time.com/5216601/vladimir-putin-visits-kemerovo-shopping-mall-fire}}. With respect to evacuation in public buildings, the researches is already substantial. Models of pedestrian evacuation dynamics span multiple scales, including macroscopic level (network modeling), mesoscopic level (continuous flow modeling), microscopic level (individual motion modeling), etc. Given the pronounced heterogeneity among pedestrians, realistic evacuation analyses require an explicit representation of decision behavior to achieve high-fidelity simulations. Accordingly, the literature primarily examines tactical behavior, i.e., route choice in evacuation contexts.

To this end, a substantial body of research spans modeling approaches, field observations, and integrative studies. Representative strands include microscopic navigation modeling \citep{kretz2014user}; the two-level model integrating the operational and tactical behaviors \citep{zhang2016simulation, yu2020consideration}; evacuation simulation that leverage visual information \citep{yu2024consideration, zhou2021route}; information-sharing navigation modeling (leader–follower behavior) \citep{liu2018crowd}; analyses of herd behavior \citep{haghani2019imitative}; experimental observation and modeling of route choice \citep{crociani2016route}, including the route choice of wheelchair users \citep{feliciani2020efficiently}; applications of cumulative prospect theory \citep{gao2022integration} and boundedly rational decision \citep{wang2022modeling} in route choice; investigations based on virtual reality \citep{lovreglio2016mixed, feng2021using, zhiming2021study}; experimental investigations \citep{jin2025findings}; and macroscopic statistics on pedestrian route-choice preferences \citep{sevtsuk2021big}, etc. Narrowing the focus to pedestrian evacuation, most studies that incorporate individual heterogeneity concentrate on decision making at the microscopic and mesoscopic scales. By contrast, research on macroscopic route choice remains limited, even though this is precisely where many practical challenges arise. From a macroscopic standpoint, evacuation can be formulated as an optimization problem, as illustrated by representative studies on multi-objective route design for heterogeneous evacuees \citep{yang2023multi}, dynamic pedestrian assignment on large networks \citep{lilasathapornkit2022dynamic}, macroscopic flow modeling \citep{goatin2009macroscopic}, and network-based optimization frameworks \citep{tang2025flexible}. Queueing theoretic models have also been discussed as an alternative for pedestrian traffic systems \citep{mitchell2001topological}. However, approaches that rely on steady state performance (typically requiring traffic intensity \(\rho<1\)) are ill suited to evacuation contexts. During evacuation, arrivals are highly concentrated over a short horizon with \(\rho \gg 1\), then decay to zero permanently; moreover, the arrival process is time varying and generally non Markovian. These characteristics undermine the steady-state assumptions underlying standard queueing approximations, thus limiting their applicability to evacuation problems.

As the research scale expands and pedestrian behavior becomes highly stochastic, it is common to posit an independent and identically distributed (IID) assumption, whereby average behavior is characterized through statistical inference. This perspective has long been central to traffic assignment research. Within this context, Braess' paradox has become a canonical phenomenon, highlighting how changes in network design can produce counterintuitive effects \citep{braess1968paradoxon,braess2005paradox}. Regarding prevalence, both theoretical analyses and empirical investigations suggest that Braess' paradox can arise under plausible conditions and is not merely a theoretical curiosity \citep{valiant2006braess,youn2008price,ccolak2016understanding}. Empirical observations also point to paradoxical outcomes, such as the temporary closure of 42nd Street in Manhattan for Earth Day in 1990, which reduced congestion in the surrounding area \citep{gina1990if}.  Under reasonable assumptions, it may occur with nontrivial frequency \citep{di2014braess}, and in some conditions its occurrence is about as likely as its nonoccurrence \citep{steinberg1983prevalence}, often within specific ranges of travel demand \citep{pas1997braess}. A closely related insight is Smith' paradox, which shows that increasing delay locally can reduce delay globally \citep{smith1978road}. Related paradoxes have been reported beyond transportation, including mesoscopic electron networks where adding a conductive path lowered overall conductance \citep{pala2012transport} and analogous effects in mechanical and electrical networks \citep{cohen1991paradoxical}.

Compared to Braess’ paradox in daily traffic contexts \citep{ccolak2016understanding}, the paradox is often discussed in evacuation scenarios because travel demand is typically static rather than dynamic \citep{lilasathapornkit2022dynamic}.  In this line of work, researchers studies of traffic-control measures during emergency evacuations \citep{huibregtse2012blocking, kimms2016network}; network design to anticipate selfish evacuation routing \citep{kimms2016network} and identify links with negative utility by comparing user equilibrium (UE) and system optimal (SO) configurations \citep{huibregtse2012blocking}. Despite the sustained attention it has received in road traffic, rigorous treatments in pedestrian systems remain scarce \citep{crociani2016multidestination}. In evacuation contexts, studies frequently discuss how appropriately placed obstacles can facilitate pedestrian egress \citep{cristiani2017handling}; however, these investigations are typically oriented toward physical mechanisms rather than network optimization, and thus are not directly situated within the network assignment framework of Braess' paradox. Although some simulation-based examples exist \citep{crociani2016multidestination}, to the best of our knowledge, Braess’ paradox has never been observed empirically in pedestrian traffic systems. Pedestrian route choice is a highly complex process, influenced by factors such as the completeness of information and the timeliness of decision-making. These factors result in significant discrepancies between real-world observations and idealized scenarios \citep{barreiro2022stochastic}. Pedestrian evacuation traffic appears to be inherently well suited to planning problems. In emergency settings, evacuees exhibit intrinsic autonomy in decision making; moreover, demand loading is concentrated, which makes a static traffic assignment formulation appropriate. To this end, we combine evidence from supervised experiments with the static traffic assignment framework to investigate Braess’ paradox in pedestrian evacuation systems.

The main contributions of this study are as follows: (1) We investigate the existence of Braess’ paradox in pedestrian evacuation traffic. The presence of Braess’ paradox in pedestrian evacuation networks holds significant implications for the design and safety management of large-scale public infrastructure, yet it has long been overlooked; (2) We design a refined experiment for the calibration of parameters related to pedestrian route choice behavior. This experimental geometry allows for flexible editing of the geometric configuration, thereby enabling broader explorations of pedestrian route choice problems; (3) By adopting a stochastic user equilibrium (SUE) framework with multiplicative errors, we demonstrate how the calibration results obtained from small-scale experiments can be transferred to large-scale network instances. In addition, we present equivalent convex optimization formulations of both path-flow and link-flow assignments with multiplicative errors.

The subsequent sections of the paper are organized as follows: in Section~\ref{section2}, the theoretical foundations of the UE and SUE models are presented.  
In particular, an SUE model based on multiplicative error terms is formulated, and the corresponding link-flow and path-flow programs are derived.  
In Section~\ref{section3}, the parameters of both UE and SUE models are calibrated using the experimental results, and the comparative analysis demonstrates that the model outcomes exhibit a high degree of consistency with the empirical data.  
The case study is reported in Section~\ref{section4}, where the Braess paradox phenomenon is explored from a broader perspective.  
Finally, Section~\ref{section5} provides a summary of the conclusions drawn from the study.

\section{Model Properties}\label{section2}

We first introduce the fundamental theory of static traffic assignment and apply it to evacuation modeling. We represent the evacuation network as a directed graph \(\mathcal{G}=(\mathcal{V},\mathcal{A})\), where \(\mathcal{V}\) is the set of vertices (nodes) and \(\mathcal{A}\subseteq \mathcal{V}\times\mathcal{V}\) is the set of directed links (arcs). For any link \(a\in\mathcal{A}\), let \(x_a\ge 0\) denote the link flow, and let \(t_a(x_a)\) denote the associated evacuation cost or evacuation time. For a given origin--destination (OD) pair \((r,s)\) with demand \(d_{rs}\), let \(\mathcal{K}_{rs}\) denote the set of admissible paths from \(r\) to \(s\). Let $\mathcal{X}$ denote the set of link flows subject to network flow conservation and reachability constraints, with $\mathbf{x}\in\mathcal{X}$ representing a feasible flow allocation. The cost and flow of path \(i\in\mathcal{K}_{rs}\) under the link-flow vector \(\mathbf{x}=\{x_a\}_{a\in\mathcal{A}}\) is written as \(\tau_i(\mathbf{x})\) and \(f_i^{rs}\), where $\delta_{a,i}^{rs}\in{0,1}$ denotes the link–path incidence indicator. With the notation, we first examine the basic properties of UE and SUE.

\subsection{UE and SO}\label{subsection2.1}
Building on Wardrop's first and second principles \citep{wardrop1952road}, the UE and SO flow can be obtained as the solution to the following programs:

\begin{equation}
\begin{aligned}
\min_{\mathbf{x}\in\mathcal{X}} \; & Z_{\mathrm{UE}} = \sum_{a \in \mathcal{A}} \int_0^{x_a} t_a(u)\, du \\
\text{s.t.} \quad 
& \sum_{k \in \mathcal{K}_{rs}} f_k^{rs} = d_{rs}, \quad \forall (r,s) \\
& x_a = \sum_{(r,s)} \sum_{k \in \mathcal{K}_{rs}} \delta_{a k}^{rs} f_k^{rs}, \quad \forall a \in \mathcal{A} \\
& f_k^{rs} \geq 0, \quad \forall k,(r,s)
\end{aligned}
\label{1}
\end{equation}

\begin{equation}
\begin{aligned}
\min_{\mathbf{x}\in\mathcal{X}} \; & Z_{\mathrm{SO}} = \sum_{a \in \mathcal{A}} x_a \cdot t_a(x_a)\\
\text{s.t.} \quad 
& \sum_{k \in \mathcal{K}_{rs}} f_k^{rs} = d_{rs}, \quad \forall (r,s) \\
& x_a = \sum_{(r,s)} \sum_{k \in \mathcal{K}_{rs}} \delta_{a k}^{rs} f_k^{rs}, \quad \forall a \in \mathcal{A} \\
& f_k^{rs} \geq 0, \quad \forall k,(r,s)
\end{aligned}
\label{2}
\end{equation}

Eq.\eqref{1} states the equilibrium condition under which users minimize their own evacuation time, that is, the UE flow allocation, and the Eq.\eqref{2} states the centralized optimization that minimizes total evacuation time, namely the SO flow allocation. When \(t_a(x_a)\) is convex and nondecreasing, the both corresponding programs are convex. With the given information above, we can get sufficient condition for the equivalence between UE and SO solutions. It holds under the following condition: 

\textit{If the cost function $t_a(x_a)$ for each link $a$ in the network is a monomial of the same order in flow, i.e., $t_a(x_a) = k_a x_a^n, \quad \text{with } k_a > 0,\, n > 0$, then the UE and the SO traffic assignment problems admit the same flow distribution as their solution.}

This classic property indicates that, in network flow analysis using a link performance function of the Bureau of Public Roads (BPR) form ($\beta>0$), the UE flow pattern approaches the SO flow pattern as demand increases.

\subsection{SUE with Multiplicative Error}\label{subsection2.2}

Our perspective is that, when pedestrians evacuate across different spatial scales, the magnitudes of both the error term and the observed quantities vary proportionally. Accordingly, employing a multinomial logit model with multiplicative rather than additive error allows evacuation behavior to be analyzed across different scales within a unified modeling framework. In this formulation, the perceived (random) cost is assumed to follow a multiplicative structure:

\begin{equation}
C_i=\tau_i(\mathbf{x})\,\varepsilon_\textit{i},\qquad \textit{i}\in\mathcal{K}_{rs},
\label{3}
\end{equation}

where $\{\varepsilon_i\}$ are i.i.d.\ across alternatives and individuals with $\varepsilon_i>0$. 
Because the logarithm is strictly increasing on $\mathbb{R}^{+}$, the choice rule

\begin{equation}
P\left(i\,\big|\,\mathbf{x},rs\right)=
p\left(-\ln\tau_i(\mathbf{x})-\ln\varepsilon_i\ \ge\ -\ln\tau_j(\mathbf{x})-\ln\varepsilon_j,\ \forall j\in\mathcal{K}_{rs}\right).
\label{4}
\end{equation}

Define the reparameterization $-\ln\varepsilon_i=\xi_i/\theta$ with scale parameter $\theta>0$, where $\{\xi_i\}$ are i.i.d.\ random variables with common distribution and constant scale across $i$. The model admits an additive random-utility representation with a logarithmic systematic component:

\begin{equation}
\bar U_i\;=\;-\theta\,\ln\tau_i(\mathbf{x})+\xi_i,\qquad i\in\mathcal{K}_{rs}.
\label{5}
\end{equation}

Assume the error terms $\xi_i$ are i.i.d.\ Gumbel$(0,1)$, the multinomial logit (MNL) probabilities are \citep{fosgerau2009discrete}:

\begin{equation}
p_i^{rs}(\mathbf{x})\;=\;P\!\left(i\,\big|\,\mathbf{x},rs\right)
=\frac{\exp\!\big(-\theta\,\ln\tau_i(\mathbf{x})\big)}
{\displaystyle\sum_{j\in\mathcal{K}_{rs}}\exp\!\big(-\theta\,\ln\tau_j(\mathbf{x})\big)}
=\frac{\tau_i(\mathbf{x})^{-\theta}}
{\displaystyle\sum_{j\in\mathcal{K}_{rs}}\tau_j(\mathbf{x})^{-\theta}},
\qquad \theta>0.
\label{6}
\end{equation}

Let $a$ index directed links and $rs$ index OD pairs with fixed demand $d_{rs}>0$, the path cost is
\begin{equation}
\tau_i(\mathbf{x})=\sum_{a\in \mathcal{A}}\delta_{a,i}^{rs}\,t_a(x_a)>0.
\label{7}
\end{equation}

Define the Beckmann potential (UE): $\sum_{a\in \mathcal{A}}\int_{0}^{x_a} t_a(u)\,du$. The SUE of multiplicative error is equivalently characterized by the following link flow program (convex under the condition that $\ln \tau_i(\mathbf{x})$ is concave) \citep{akamatsu1997decomposition}:

\begin{equation}
\min_{\mathbf{x}\in\mathcal{X}}\;
\sum_{a\in \mathcal{A}}\int_{0}^{x_a} t_a(u)\,du+\sum_{rs}\frac{d_{rs}}{\theta}\,
\ln\!\left(\sum_{i\in\mathcal{K}_{rs}} \tau_i(\mathbf{x})^{-\theta}\right)
\label{8}
\end{equation}

For any fixed $\mathbf{x}$ and OD pair $(r,s)$, the following identity holds (convex conjugate)\citep{fisk1980some}:

\begin{equation}
\frac{d_{rs}}{\theta}\,\ln \sum_{i\in\mathcal{K}_{rs}}\tau_i(\mathbf{x})^{-\theta}
=\max_{\{f_i^{rs}\ge 0,\;\sum_i f_i^{rs}=d_{rs}\}}
\left[-\sum_{i\in\mathcal{K}_{rs}} f_i^{rs}\,\ln \tau_i(\mathbf{x})
-\frac{1}{\theta}\sum_{i\in\mathcal{K}_{rs}} f_i^{rs}\,\ln f_i^{rs}
\right]
+\frac{d_{rs}}{\theta}\ln d_{rs}.
\label{9}
\end{equation}

The term $\frac{d_{rs}}{\theta}\ln d_{rs}$ is a constant term and can be dropped. Using \ref{9}, the link-based program \ref{8} can be written as the following equivalent two-level program with path flows:

\begin{equation}
\min_{\mathbf{x}\in\mathcal{X}}\left\{
\sum_{a\in \mathcal{A}}\int_{0}^{x_a} t_a(u)\,du-\sum_{rs}\,
\min_{\{f_i^{rs}\ge 0,\;\sum_i f_i^{rs}=d_{rs}\}}\left[\sum_{i\in\mathcal{K}_{rs}} f_i^{rs}\,\ln \tau_i(\mathbf{x})
+\frac{1}{\theta}\sum_{i\in\mathcal{K}_{rs}} f_i^{rs}\,\ln f_i^{rs}
\right]
\right\}.
\label{10}
\end{equation}

This OD-wise inner problem in $f_i^{rs}$ has Lagrangian

\begin{equation}
\mathcal{L}^{\min}_{rs}(f,\lambda)
=\sum_{i} f_i^{rs}\,\ln \tau_i(\mathbf{x})
+\frac{1}{\theta}\sum_{i} f_i^{rs}\,\ln f_i^{rs}
+\lambda\left(\sum_{i}f_i^{rs}-d_{rs}\right).
\label{11}
\end{equation}

Stationarity for any optimal $f_i^{rs}>0$ gives:

\begin{equation}
f_i^{rs}=e^{-1-\theta\lambda}\,\tau_i(\mathbf{x})^{-\theta}
\label{12}
\end{equation}

Enforcing the OD conservation $\sum_{i\in\mathcal{K}_{rs}} f_i^{rs}=d_{rs}$ gives:

\begin{equation}
e^{-1-\theta\lambda}\;=\; \frac{d_{rs}}{\sum_{j\in\mathcal{K}_{rs}} \tau_j(\mathbf{x})^{-\theta}}.
\label{13}
\end{equation}

Substituting back yields the multiplicative error path–flow expression: 
\begin{equation}
f_i^{rs}=\frac{d_{rs}\,\tau_i(\mathbf{x})^{-\theta}}{\sum_{j\in\mathcal{K}_{rs}}\tau_j(\mathbf{x})^{-\theta}}
\label{14}
\end{equation}

\begin{figure}[ht!]
\centering
\includegraphics[scale=0.6]{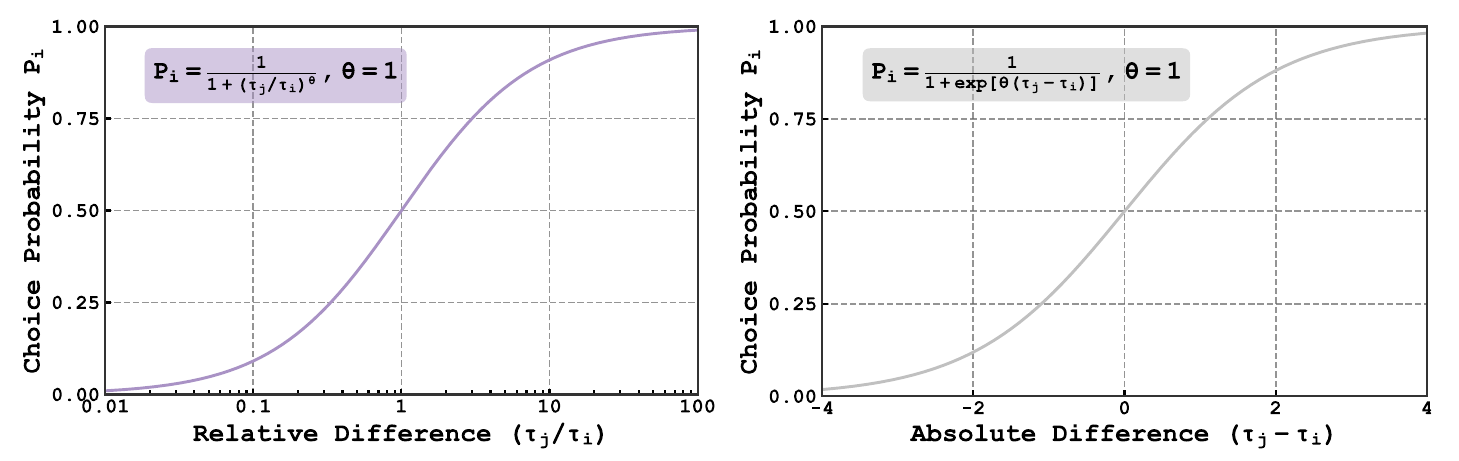}
\caption{Comparison of choice probabilities in binary logit models based on (a) multiplicative and (b) additive errors.}
\label{fig1}
\end{figure}

Accordingly, the flow assignment follows the log–logistic choice rule. By substituting Eq.\eqref{14} into Eq.\eqref{10}, we obtain the link flow program (Eq.\eqref{8}). This modification transforms path flow allocation in the classical logit SUE framework from being based on absolute errors to being based on relative errors. Taking the simplest case of a binary logit choice as an example, a comparison of the resulting choice probabilities is illustrated in Fig.\ref{fig1}.

\section{Experimental Investigation} \label{section3}

\begin{figure}[ht!]
\centering
\includegraphics[scale=0.6]{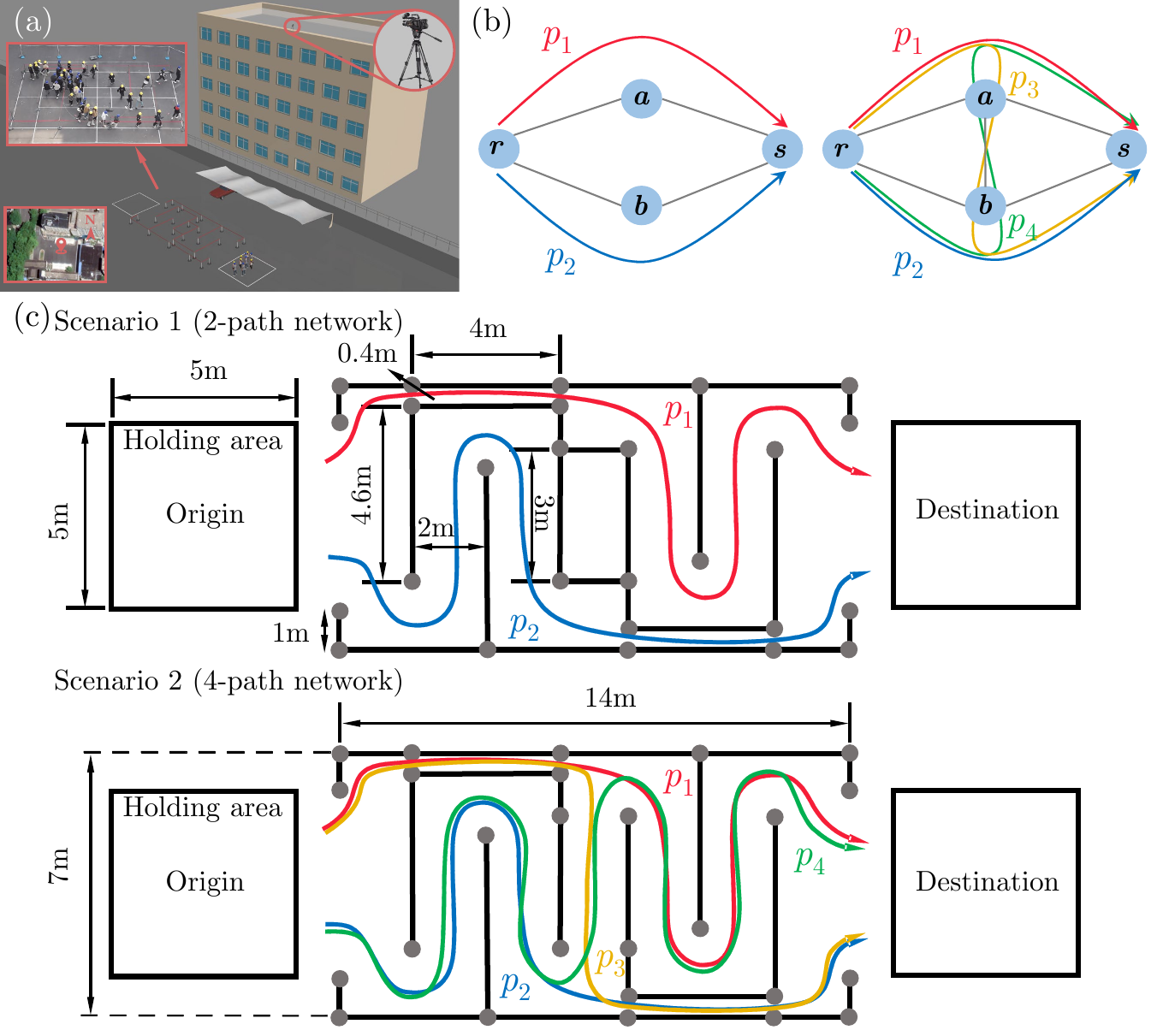}
\caption{Route choice experiment and setup, (a) experimental schematic (location: 30$^\circ$30'59"N 114$^\circ$20'58"E), (b) corresponding topology network, and (c) geometric configurations of the experimental scenario.}
\label{fig2}
\end{figure}

To investigate the potential occurrence of Braess' paradox in pedestrian evacuation systems, and inspired by \citet{crociani2016multidestination}, we designed an experimental scenario based on the Braess 5-link network topology, as illustrated in Fig.\ref{fig2}. In Scenario 1(S1) and Scenario 2 (S2), which were designed as 2-path and 4-path networks, respectively, each experimental trial was conducted with evacuation populations increasing in increments of 10 participants, up to a maximum of 60 individuals. The gender ratio of participants was maintained at 1:1 in all cases. Details concerning the implementation of the experiment and the specific results of the evacuation data are provided in Appx.\ref{Appendix A}.

\subsection{Parameter Calibration}
\label{subsection3.1}

Based on the experimental results, we first perform the parameter calibration of the UE and SUE models. Compared with road traffic, pedestrian traffic systems exhibit greater robustness, as they are not sensitive to queue spillback. In other words, even under large demand conditions, pedestrian traffic systems are unlikely to experience traffic breakdown, a phenomenon that frequently occurs in road networks. Therefore, we did not adopt the general form of the BPR function (characterized by a concave formulation with $\beta > 1$) as the pedestrian link performance function. We first consider an evacuating crowd organized in a single queue characterized by a constant headway $t_h$. In this setting, the evacuation time of the $x$-th pedestrian can be expressed as $t = t_{0} + x \cdot t_{h}$, where $x \in \mathbb{N}$. The free-flow evacuation time $t_0$ is given by $\frac{\ell}{v_{\mathrm{max}}}$, with $v_{\mathrm{max}}$ set as 4 m/s denoting the maximum evacuation speed and $\ell$ denoting the corridor length. Furthermore, the corresponding link performance function associated with a single queue can be derived as follows:

\begin{equation}
t_a(x_a) = \frac{{\int_0^{{x_a}} {\left( {{t_0} + u\cdot{t_h}} \right)du} }}{{{x_a}}} = {t_0}  + \frac{x_a \cdot t_h}{2}, x_a\in {\mathbb{N}}
\label{15}
\end{equation}

In the context of evacuation within public spaces, we assume that pedestrians move in a queue formation along corridors, where the number of such queues is positively correlated with the corridor width, thereby yielding the latency function of movement within the corridor charactered by length $\ell$ and width $w$:

\begin{equation}
t_a(x_a) = t_0 + \frac{x_a \cdot t_h \cdot W}{2 \cdot w}, x_a\in {\mathbb{N}}
\label{16}
\end{equation}

Here, $W$ denotes the equivalent single-queue width. It can be observed that the link performance function is expressed as a linear function of the corridor length $\ell$ and the corridor width $w$, which can be interpreted as a special case of the BPR function with $\beta = 1$. Such a specification (a linear increase in evacuation time with the number of evacuees under fixed configurations) is in fact highly consistent with empirical investigations \citep{schadschneider2008evacuation}. Besides, in our evacuation network composed of corridors, we consider symmetric link travel times, meaning that whether a edge is configured for uni-directional or bi-directional flow share the same latency characteristics. Experimental results have highlighted the differences in fundamental diagram characteristics between uni and bi-directional configurations \citep{zhang2012ordering}. Due to the lane-formation mechanism, these differences do not result in significant variations in latency characteristics.

\begin{figure}[ht!]
\centering
\includegraphics[scale=0.5]{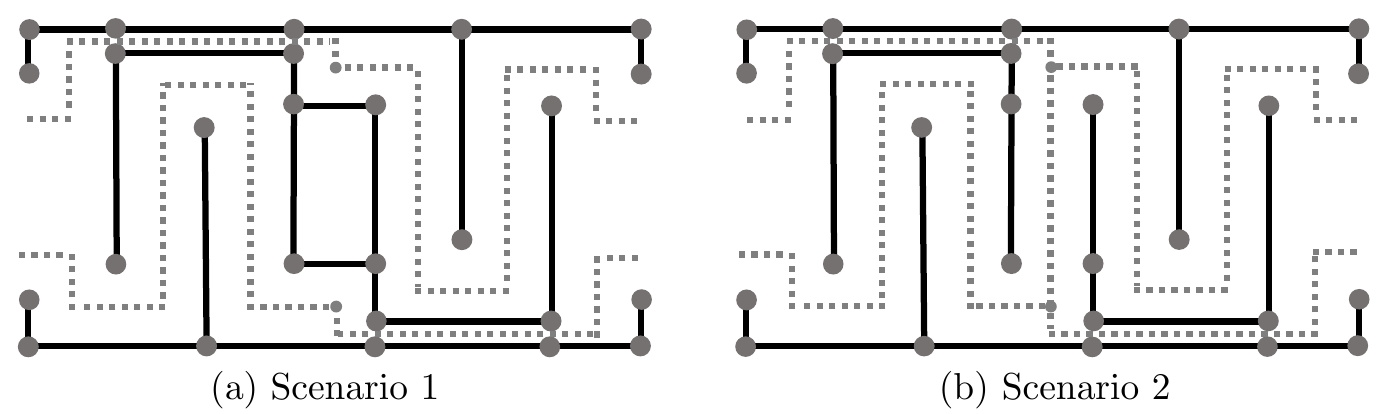}
\caption{Illustration of the geometric configurations in Scenario 1 (a) and Scenario 2 (b), with dashed lines indicating the centerline of corridors.}
\label{fig3}
\end{figure}

Based on the above settings, it is first necessary to understand the geometric configuration of the experiment, as illustrated in Fig.\ref{3}. According to the centerline of the corridors and their widths, the geometric parameters of the edges (an edge is represented by two directed arcs in opposite directions) in the topological network can be specified as follows: $\ell_{ra} = \ell_{bs} = 9.6\,\text{m}$, $\; w_{ra} = w_{bs} = 0.4\,\text{m}$, $\; \ell_{rb} = \ell_{as} = 17.2\,\text{m}$, $\; w_{rb} = w_{as} = 2\,\text{m}$, $\; \ell_{ab} = 5\,\text{m}$, and $\; w_{ab} = 2\,\text{m}$. Based on these parameters, we proceed with the calibration of the UE and SUE models accordingly.  
We adopt the root mean square error (RMSE) of the average evacuation time as the indicator of result consistency, which is formulated as:

\begin{equation}
\mathrm{RMSE}=\sqrt{\frac{1}{s\cdot n}\sum\limits_{s,n}\bigl(t_{\exp}-t_{\text{mod}}\bigr)^{2}}.
\label{17}
\end{equation}

\begin{figure}[ht!]
\centering
\includegraphics[scale=0.7]{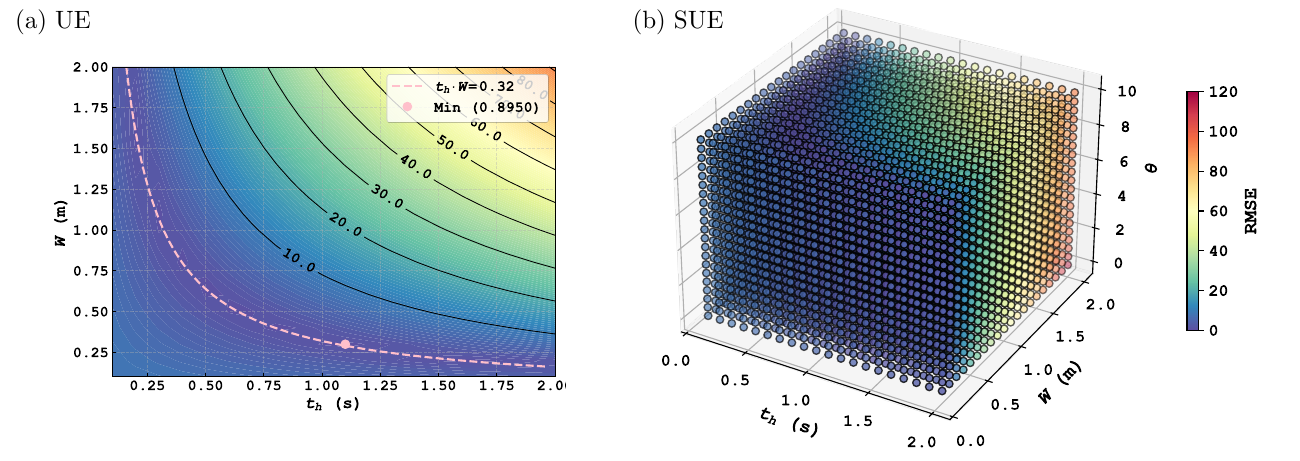}
\caption{Distribution of RMSE under different parameter configurations: (a) UE and (b) SUE.}
\label{fig4}
\end{figure}

A parameter calibration was conducted by comparing the results of the UE and SUE models with the experimental results, as illustrated in Fig.\ref{fig4}. In this process, $t_h$ was varied within the span $[0.1,\,2]$ with an increment of $0.1\,\text{s}$, $W$ was varied within the span $[0.1,\,2]$ with an increment of $0.1\,\text{m}$, and the scale parameter $\theta$ was varied within the span $[0.5,\,10]$ with an increment of $0.5$. Based on the parameter calibration results, the optimal configuration for the UE model was obtained as $t_h = 1.1$ and $W = 0.3$, corresponding to $\mathrm{RMSE} = 0.8950$s. For the SUE model, the optimal configuration was identified as $t_h = 0.8$, $W = 0.4$, and $\theta = 4.5$, with a corresponding $\mathrm{RMSE} = 0.7936$s. For the purpose of comparison, the parameters throughout the research were uniformly set to $t_h = 0.8$, $W = 0.4$, and $\theta = 4.5$, under which the RMSE of the UE model was calculated as $0.9007$s.

Given the calibrated parameters, a comparison was made between the experimental data and the results of the UE and SUE models, focusing on the average evacuation time and the path ratios, as illustrated in Fig.\ref{5}. The figure indicates that the SUE model demonstrates closer agreement with the experimental results in terms of average evacuation time, whereas the UE model provides more accurate estimates for the path ratios (S2 group).

\begin{figure}[ht!]
\centering
\includegraphics[scale=0.8]{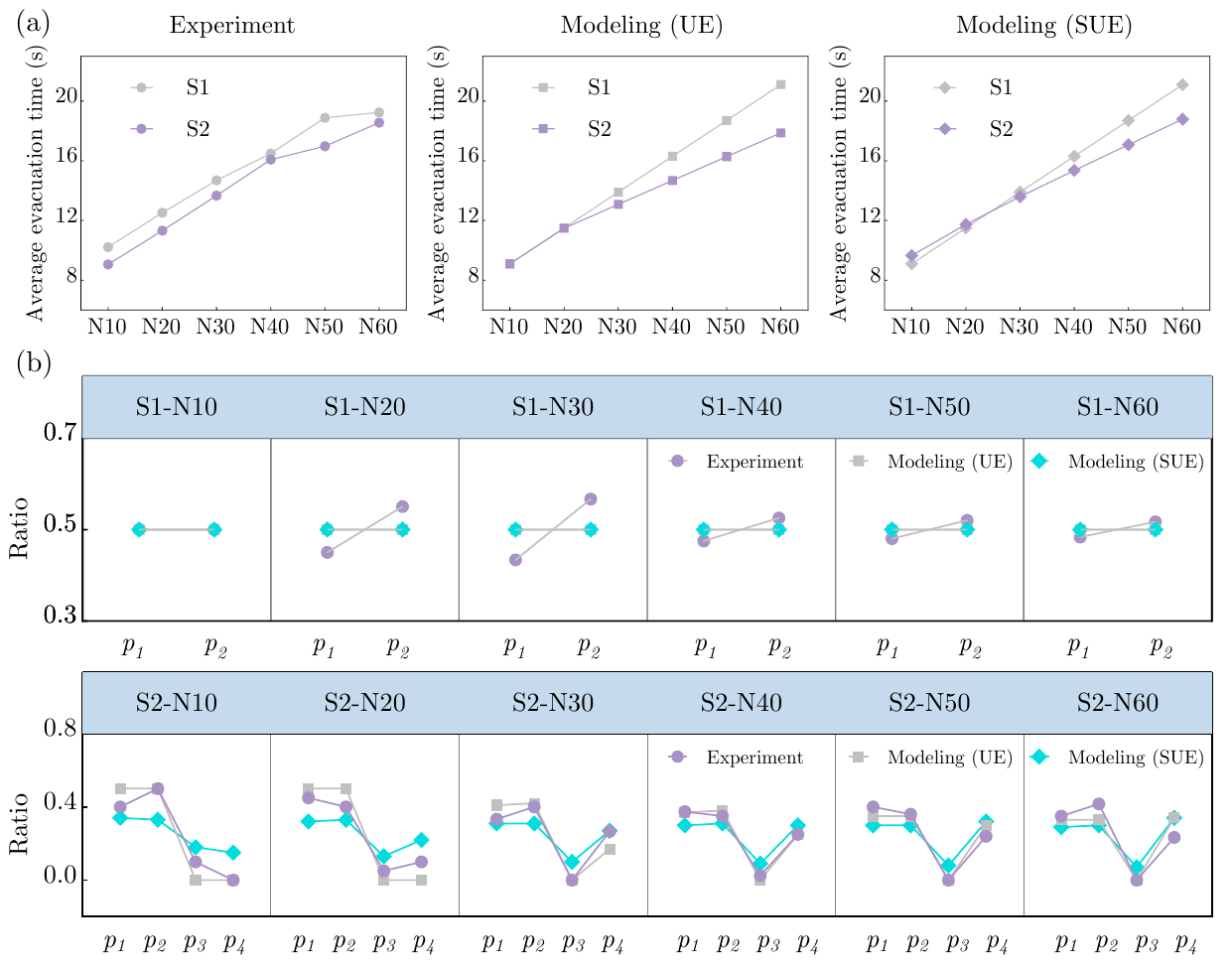}
\caption{Comparison of experimental results and model outputs of UE and SUE: (a) evacuation time; (b) path ratios.}
\label{fig5}
\end{figure}

\begin{figure}[ht!]
\centering
\includegraphics[scale=0.8]{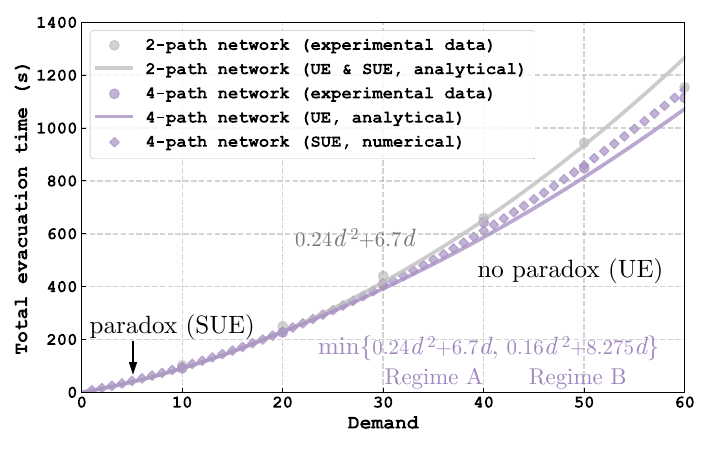}
\caption{Trend of total evacuation time of experimental results and model results of UE and SUE.}
\label{fig6}
\end{figure}

Based on the experimental and model data, the variation trend of the total evacuation time under both configurations was obtained, as illustrated in Fig.\ref{fig6}. From the results of the experiment and the UE model, no occurrence of Braess' paradox was observed. In contrast, within the SUE model results, under a certain level of evacuation demand (about $d_{rs}<20$, see Fig.\ref{fig5} (a)), the total evacuation time of the 4-path network exceeded that of the 2-path network, indicating the manifestation of Braess' paradox. This suggests that, considering evacuees' limited knowledge of path information, Braess' paradox may potentially arise.

\subsection{Possibility of Braess’ Paradox under Varying Geometry} \label{subsection3.2}

Based on the experimental geometry configuration, it was observed that Braess’ paradox arose under limited knowledge of path information (SUE), whereas no paradox occurred under complete knowledge of path information (UE). Building on the analytical tractability of UE in simple network topologies, further examination was undertaken to determine whether adjustments to the experimental geometry could substantiate the occurrence of Braess’ paradox in the context of UE assignment. We continue to analyze the symmetric 5-edge network, parameterizing each edge’s length and width as follows:
$\ell_{ra}=\ell_{bs}=\ell_{ab}=\ell_1,\;
\ell_{rb}=\ell_{as}=\ell_2,\;
w_{ra}=w_{bs}=w_1,\;
w_{rb}=w_{as}=w_{ab}=w_2$, with $\ell_1,\ell_2>0$ and $w_1,w_2>0$. The link performance function is still given by Eq.\eqref{16}, with $v_{\mathrm{max}}=4,\mathrm{m/s}$, $t_h=0.8\mathrm{s}$, and $w=0.4\mathrm{m}$.

\textbf{Scenario 1} (without edge $a$--$b$):  
Removing the edge $a$--$b$ leaves two parallel routes. By symmetry, one obtains $f_1=f_2=d_{rs}/2$. The average travel cost is then
\begin{equation}
\bar \tau_{S1}(d_{rs})
=\frac{\ell_1+\ell_2}{4}
+0.08\!\left(\frac{1}{w_1}+\frac{1}{w_2}\right)d_{rs}.
\label{18}
\end{equation}

\textbf{Scenario 2} (network with edge $a$--$b$):  
At a Wardrop UE, all used paths must have equal cost. Under the new network specification, only two regimes arise.

\textbf{Regime A} (2-path use, $p_1,p_2$): For sufficiently small demand, the central path $p_4$ is not utilized. Hence,

\begin{equation}
\bar \tau_{S2}^A(d_{rs})=\bar \tau_{S1}(d_{rs}),
\quad\text{for }0\le d_{rs}\le d_0,
\label{19}
\end{equation}

where the activation threshold $d_0$ is obtained from the condition $\tau_4=\tau_1$ under the two-path flows ($f_1=f_2=d_{rs}/2$, $f_4=0$):
\begin{equation}
d_0
=\frac{\ell_2\,w_1w_2}{0.32\,(\,w_2-w_1\,)}\,.
\label{20}
\end{equation}

Thus $d_0>0$ if and only if $w_2>w_1$. If $w_2\le w_1$, then $d_0\le0$ and the central path $p_4$ never activates.

\textbf{Regime B} (3-path use, $p_1,p_2,p_4$): For $d_{rs}\ge d_0$ (with $w_2>w_1$), the equilibrium average cost, common to all active paths, is

\begin{equation}
\bar \tau_{S2}^B(d_{rs})
=\frac{\ell_1+\ell_2}{4}
+\frac{\ell_2}{4}\cdot
\frac{\tfrac{1}{w_1}-\tfrac{1}{w_2}}{\tfrac{1}{w_1}+\tfrac{3}{w_2}}
+\frac{0.16\,(w_1+3w_2)}{\,w_2\,(w_2+3w_1)}\,d_{rs},
\qquad d_{rs}\ge d_0.
\label{21}
\end{equation}

Note that $\bar \tau_{S2}^B(d_0)=\bar \tau_{S2}^A(d_0)=\bar \tau_{S1}(d_0)$, ensuring continuity of $\bar \tau_{S2}$.

In summary,
\begin{equation}
\bar \tau_{S2}(d_{rs})=
\begin{cases}
\bar \tau_{S1}(d_{rs}),
& \text{if }w_2\le w_1,\\[8pt]
\bar \tau_{S1}(d_{rs}),
& 0\le d_{rs}\le d_0,\ \ w_2>w_1,\\[6pt]
\displaystyle
\frac{\ell_1+\ell_2}{4}
+\frac{\ell_2}{4}\cdot
\frac{\tfrac{1}{w_1}-\tfrac{1}{w_2}}{\tfrac{1}{w_1}+\tfrac{3}{w_2}}
+\frac{0.16\,(w_1+3w_2)}{\,w_2\,(w_2+3w_1)}\,d_{rs},
& d_{rs}\ge d_0,\ \ w_2>w_1.
\end{cases}
\label{22}
\end{equation}

Define the difference $\Delta(d_{rs})=\bar \tau_{S2}(d_{rs})-\bar \tau_{S1}(d_{rs})$, we can get  

Regime A ($0\le d_{rs}\le d_0$):  
\(\Delta(d_{rs})\equiv0\).  

Regime B ($d_{rs}\ge d_0$, with $w_2>w_1$):  
Subtracting Eq.\eqref{21} from Eq.\eqref{18} yields a linear expression:
\begin{equation}
\Delta(d_{rs})
=\frac{\ell_2}{4}\cdot
\frac{\tfrac{1}{w_1}-\tfrac{1}{w_2}}{\tfrac{1}{w_1}+\tfrac{3}{w_2}}
+\Bigg[
\frac{0.16\,(w_1+3w_2)}{\,w_2\,(w_2+3w_1)}
-0.08\!\left(\frac{1}{w_1}+\frac{1}{w_2}\right)\Bigg]\,d_{rs}.
\label{23}
\end{equation}

Its derivative simplifies to

\begin{equation}
\Delta'(d_{rs})
=-\,\frac{0.08\left(\tfrac{1}{w_1}-\tfrac{1}{w_2}\right)^2}{\tfrac{1}{w_1}+\tfrac{3}{w_2}}
\;\le\;0.
\label{24}
\end{equation}

This results implying that $\Delta(d_{rs})$ is monotonically nonincreasing on Regime~B (strictly decreasing whenever $w_1\neq w_2$). Moreover, at the activation point $d_0$, one has $\Delta(d_0)=0$, and thus $\Delta(d_{rs})<0$ for all $d_{rs}>d_0$.  

Accordingly, under the specified conditions, the UE-based analysis shows that no geometric modification leads to the emergence of Braess’ paradox:

\begin{itemize}
\item If $w_2\le w_1$, the central path never activates, and hence $\Delta(d_{rs})\equiv0$.
\item If $w_2>w_1$, then $\Delta(d_{rs})=0$ for $d_{rs}\le d_0$, while $\Delta(d_{rs})<0$ for $d_{rs}>d_0$.
\end{itemize}

Therefore, with the link performance function $t_a(x_a)=\ell/v_{\mathrm{max}}+\frac{x_a \cdot t_h \cdot W}{2 \cdot w}$, and $v_{\mathrm{max}}=4 \mathrm{m/s} ,t_h=0.8\mathrm{s},w=0.4\mathrm{m}$, the addition of the $a$--$b$ edge never increases the UE average travel cost: there is no demand range where $\Delta(d_{rs})>0$.

\section{Case Study} \label{section4}

\begin{figure}[ht!]
\centering
\includegraphics[scale=0.32]{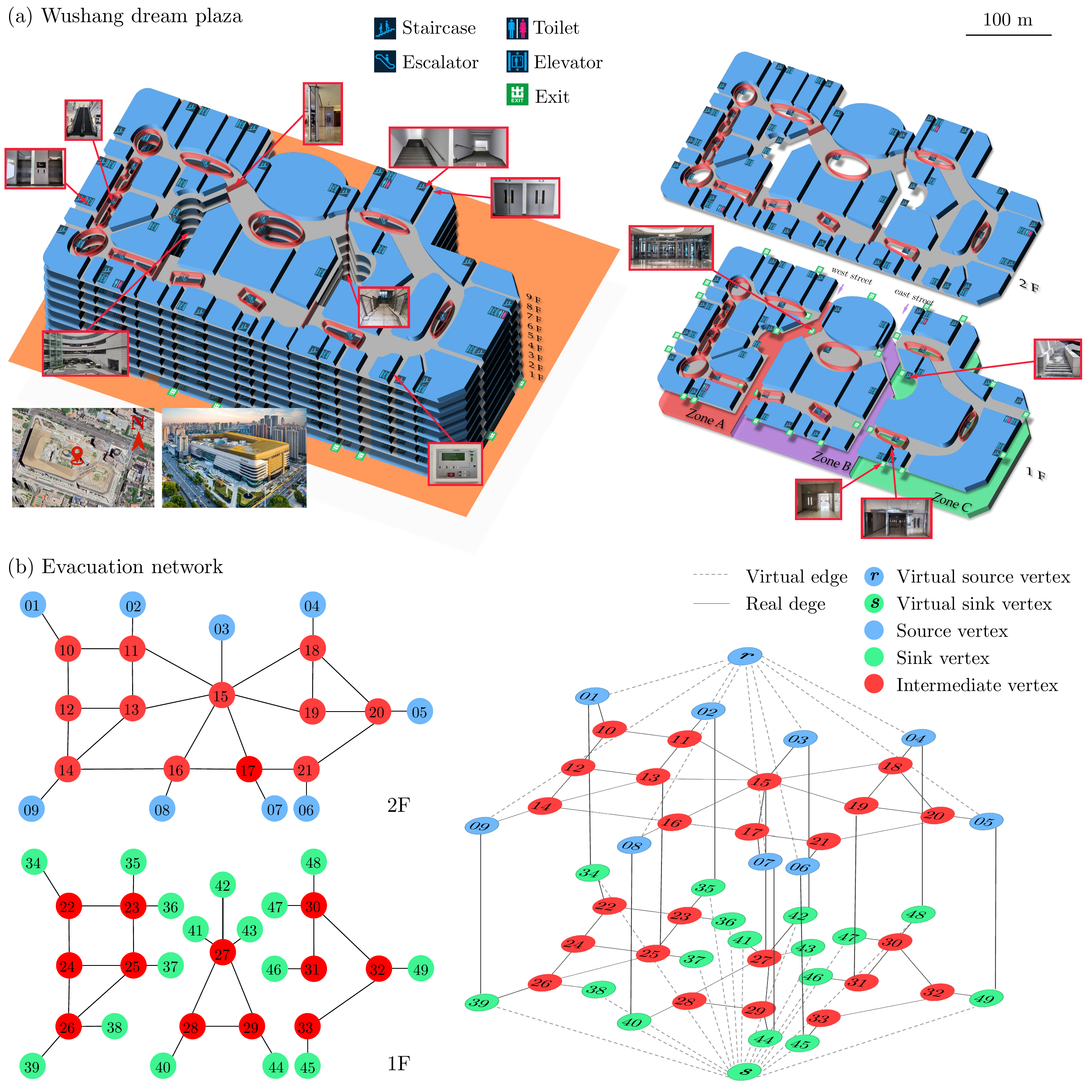}
\caption{Three-dimensional configuration and plan view of Wushang Dream Plaza (location: 30$^\circ$31'47"N 114$^\circ$20'10"E); (b) Topology network mapped from Wushang Dream Plaza.}
\label{fig7}
\end{figure}

 Based on the parameters calibrated above, Wushang Dream Plaza is employed as a case study to explore whether Braess' paradox may arise in pedestrian evacuation within large indoor public venues. Such an investigation holds significant implications for emergency management and response strategies.  Wushang Dream Plaza, which opened on November 9, 2022, is located at the intersection of Wuluo Road and Baotongsi Road in the central commercial district of Wuchang, Wuhan, adjacent to Wuhan University of Technology. The project has four underground levels, with Levels B2 to B4 used for parking, and nine floors above ground. The total gross floor area is 800,000 square meters, presented as the world’s largest purely commercial complex\footnote{\url{https://www.wuhan.gov.cn/sy/whyw/202210/t20221028\\\_2077066.shtml}}, as illustrated in Fig.\ref{fig7}. In such mega-plaza, evacuation during emergencies constitutes a critical issue. Given that the actual mall comprises nine above-ground floors and one underground floor in active use, evacuation from upper and underground levels is typically concentrated through vertical staircases, with little consideration of horizontal path alternatives. To simplify the analysis, only the planar topology of the first and second floors is considered. It is assumed that the evacuation demand from the floors above the second is concentrated at the vertical staircases, while the evacuation demand from the basement levels is disregarded. Moreover, considering the fire safety requirements under emergency conditions, evacuation by means of escalators and elevators is not considered. The corresponding evacuation topology network is illustrated in Fig.\ref{fig7}(b).

In emergency conditions, evacuation in such environments can be conceptualized as the assignment of multiple OD flows on a pedestrian traffic network. By introducing virtual source and sink nodes whose associated link latency is set to zero, the multi-OD flow problem can be reformulated as an equivalent single OD flow assignment. In this formulation, evacuation demand is aggregated at the virtual source node and ultimately routed to the virtual sink node to exit the network. To ensure the rationality of the network flow assignment, several fundamental assumptions are specified. First, only the total demand is fixed, while the inflows at each source node (vertices 01–09) and the outflows at each sink node (vertices 34–49) vary with the flow configuration. Second, source and sink nodes are excluded from being used as intermediate nodes during path generation. This condition implies that edges connected to the source and sink nodes do not carry bidirectional flows. Under these settings, an evacuation network is obtained that consists of 51 nodes (including two virtual nodes) and 93 edges (including 25 virtual edges), with a total of 8,407 feasible evacuation paths. In the computation of the SUE, only the 40 paths with the shortest evacuation times are updated at each iteration, owing to the large size of the path set. The iterative process terminates either upon reaching 5000 iterations or when the difference between successive iterations falls below 0.001 seconds.

The configurations of corridor and staircase lengths and widths for each edge are provided in Table \ref{table1}. Considering that on peak days the daily number of visitors to this shopping mall exceeds 200{,}000%
\footnote{\url{https://www.hubei.gov.cn/hbfb/rdgz/202505/t20250503\_5638935.shtml}}, 
it is implied that tens of thousands of individuals may be present simultaneously within the facility. 
Accordingly, the maximum evacuation demand is set at 50{,}000 persons.

\begin{table}[htbp]
\raggedright
\caption{Edge length and width configurations in the evacuation network.}
\label{table1}
\begin{tabular}{ccc|ccc|ccc|ccc|ccc|ccc|ccc}
\toprule
Edge & $\ell$ & $w$ &
Edge & $\ell$ & $w$ &
Edge & $\ell$ & $w$ &
Edge & $\ell$ & $w$ &
Edge & $\ell$ & $w$ &
Edge & $\ell$ & $w$ &
Edge & $\ell$ & $w$ \\
\midrule
$e_{1,10}$ & 30 & 5 & $e_{1,34}$ & 40 & 2 & $e_{2,11}$ & 30 & 5 & $e_{2,35}$ & 40 & 2 & $e_{3,15}$ & 30 & 5 & $e_{3,42}$ & 40 & 2 & $e_{4,18}$ & 30 & 5 \\
$e_{4,48}$ & 40 & 2 & $e_{5,20}$ & 30 & 5 & $e_{5,49}$ & 40 & 2 & $e_{6,21}$ & 30 & 5 & $e_{6,45}$ & 40 & 2 & $e_{7,17}$ & 30 & 5 & $e_{7,44}$ & 40 & 2 \\
$e_{8,16}$ & 30 & 5 & $e_{8,40}$ & 40 & 2 & $e_{9,14}$ & 30 & 5 & $e_{9,39}$ & 40 & 2 & $e_{10,11}$ & 80 & 10 & $e_{10,12}$ & 60 & 10 & $e_{11,13}$ & 60 & 10 \\
$e_{11,15}$ & 120 & 20 & $e_{12,13}$ & 80 & 10 & $e_{12,14}$ & 60 & 10 & $e_{13,14}$ & 120 & 10 & $e_{13,15}$ & 120 & 10 & $e_{13,25}$ & 40 & 2 & $e_{14,16}$ & 120 & 10 \\
$e_{15,16}$ & 120 & 10 & $e_{15,17}$ & 80 & 10 & $e_{15,18}$ & 120 & 20 & $e_{15,19}$ & 120 & 10 & $e_{15,27}$ & 40 & 2 & $e_{16,17}$ & 60 & 10 & $e_{17,21}$ & 80 & 10 \\
$e_{18,19}$ & 60 & 10 & $e_{18,20}$ & 120 & 10 & $e_{19,20}$ & 80 & 10 & $e_{19,31}$ & 40 & 2 & $e_{20,21}$ & 120 & 10 & $e_{22,23}$ & 80 & 10 & $e_{22,24}$ & 60 & 10 \\
$e_{22,34}$ & 30 & 5 & $e_{23,25}$ & 60 & 10 & $e_{23,35}$ & 30 & 5 & $e_{23,36}$ & 30 & 5 & $e_{24,25}$ & 80 & 10 & $e_{24,26}$ & 60 & 10 & $e_{25,26}$ & 120 & 10 \\
$e_{25,37}$ & 30 & 5 & $e_{26,38}$ & 30 & 5 & $e_{26,39}$ & 30 & 5 & $e_{27,28}$ & 120 & 10 & $e_{27,29}$ & 120 & 10 & $e_{27,41}$ & 30 & 20 & $e_{27,42}$ & 30 & 5 \\
$e_{27,43}$ & 30 & 20 & $e_{28,29}$ & 80 & 10 & $e_{28,40}$ & 30 & 5 & $e_{29,44}$ & 30 & 5 & $e_{30,31}$ & 60 & 10 & $e_{30,47}$ & 30 & 5 & $e_{30,48}$ & 30 & 5 \\
$e_{30,32}$ & 120 & 10 & $e_{31,46}$ & 30 & 5 & $e_{32,33}$ & 120 & 10 & $e_{32,49}$ & 30 & 5 & $e_{33,45}$ & 30 & 5 &  &  &  &  &  &  \\
\bottomrule
\end{tabular}
\end{table}

\subsection{Patterns of PoA} \label{subsection4.1}

In order to investigate the efficiency of UE and SUE configurations within this network, the variation of the Price of Anarchy (PoA) under different levels of evacuation demand is first analyzed. The PoA is defined as the ratio of the total cost associated with the equilibrium flow (UE or SUE) to that of the SO, thereby indicating the inefficiency resulting from decentralization \citep{papadimitriou2001algorithms}. The equation of CoE is expressed as:

\begin{equation}
\mathrm{PoA} =
\frac{\displaystyle \sum_{a \in \mathcal{A}, \;\mathbf{x}|\mathrm{UE/SUE}}
x_a \, t_a\!\bigl(x_a\bigr)}{\displaystyle \sum_{a \in \mathcal{A}, \;\mathbf{x}|\mathrm{SO}}
x_a \, t_a\!\bigl(x_a\bigr)}.
\label{25}
\end{equation}

\begin{figure}[ht!]
\centering
\includegraphics[scale=0.9]{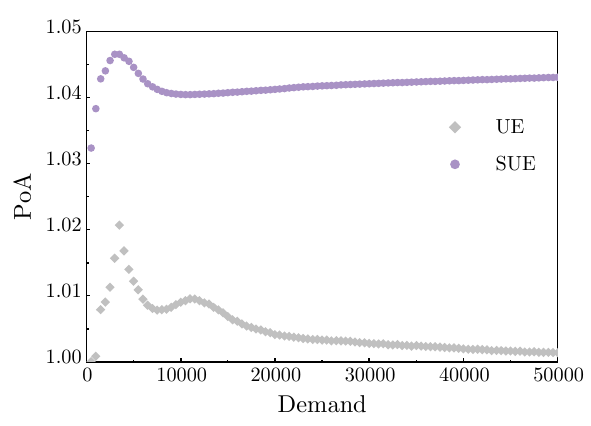}
\caption{Variation of the PoA under different demand levels in UE and SUE models.}
\label{fig8}
\end{figure}

By setting the demand increment to 500 and accumulating from 0 up to 50{,}000, the variation of the PoA curve was observed, as illustrated in Fig.\ref{fig8}. From the figure, it can be discerned that the PoA curve under the UE model exhibits a tendency of first increasing and then decreasing. This tendency is consistent with the assignment character described in Section.\ref{subsection2.1}, namely, as demand rises, the relative contribution of the fixed term (free-flow evacuation time $t_0$) diminishes, while the power-law term guides the UE flow allocation to converge toward the SO configuration, resulting in the PoA approaching unity. In contrast, the PoA curve of the SUE was observed to evolve from a fluctuating to a stable pattern, with its efficiency being significantly lower than that of the UE flow configuration.

\subsection{Flow Patterns under UE and SUE} \label{subsection4.2}

\begin{figure}[ht!]
\centering
\includegraphics[scale=0.85]{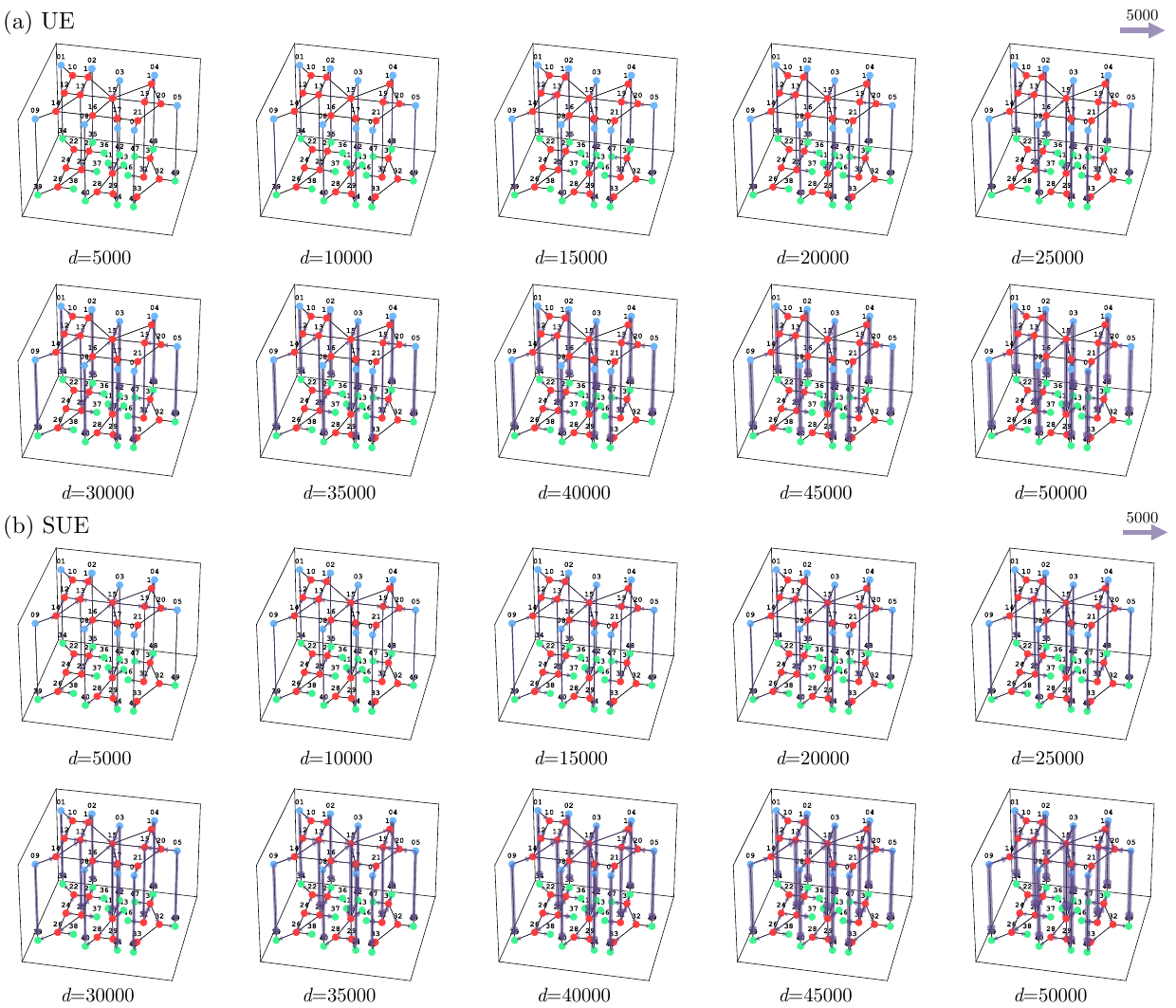}
\caption{Network flow patterns under varying evacuation demand: (a) UE configurations, (b) SUE configurations.}
\label{fig9}
\end{figure}

Furthermore, the flow configurations of the UE and SUE models were examined as demand increased in increments of 5,000 up to a maximum of 50,000, and the corresponding flow distributions were plotted, as shown in Fig.\ref{fig9}. It can be observed that, under the UE model, the allocation is highly concentrated at low demand, with evacuees primarily using the vertical evacuation staircases. As the demand increases, however, congestion along the vertical staircases intensifies, and evacuees begin to explore available horizontal routes. In particular, the flows on $e_{13,25}$, $e_{15,27}$ and $e_{19,31}$ increase significantly. In contrast, under the SUE model, pedestrians are already inclined to explore available horizontal paths even under low demand, resulting in a more balanced flow distribution. This, in turn, leads to lower evacuation efficiency compared with the UE model. At a demand level of 50,000, the flows on $e_{13,25}$, $e_{15,27}$ and $e_{19,31}$ even exceed those of the vertical staircases directly connected to the exits. This phenomenon arises due to a well-known independence of irrelevant alternatives (IIA) limitation when overlapping routes are present in the SUE model ($e_{13,25}$, $e_{15,27}$, and $e_{19,31}$ exhibit high betweenness centrality in the evacuation network).

\subsection{Sensitivity Analysis of Edges} \label{subsection4.3}

In exploring flow patterns under varying evacuation demands for both UE and SUE, the objective is to identify Braess-type edges, that is, edges whose removal results in a reduction of the total evacuation time. To achieve this objective, the criticality of edges (CoE) is employed to evaluate the isolated utility of edges within the network. The CoE is defined as the ratio of the change in total evacuation time, after the removal of a single edge under the UE or SUE flow configuration, to the total evacuation time of the original network. A positive CoE indicates that the edge has a positive utility within the network; a negative CoE implies negative utility (i.e., the occurrence of Braess’ paradox); and a CoE equal to zero indicates that the edge is unused under the given demand level. Let $\mathcal{A}^{-}$ be the set of arcs with one edge removed, the equation of CoE is expressed as:

\begin{equation}
\mathrm{CoE}\;=\;
\frac{{\displaystyle \sum_{a \in \mathcal{A}^{-}, \;\mathbf{x}|\mathrm{UE/SUE}}
x_a \, t_a\!\bigl(x_a\bigr)} - {\displaystyle \sum_{a \in \mathcal{A}, \;\mathbf{x}|\mathrm{UE/SUE}}
x_a \, t_a\!\bigl(x_a\bigr)}}
     {\displaystyle \sum_{a \in \mathcal{A}, \;\mathbf{x}|\mathrm{UE/SUE}}
x_a \, t_a\!\bigl(x_a\bigr)},
\label{26}
\end{equation}

\begin{figure}[ht!]
\centering
\includegraphics[scale=0.85]{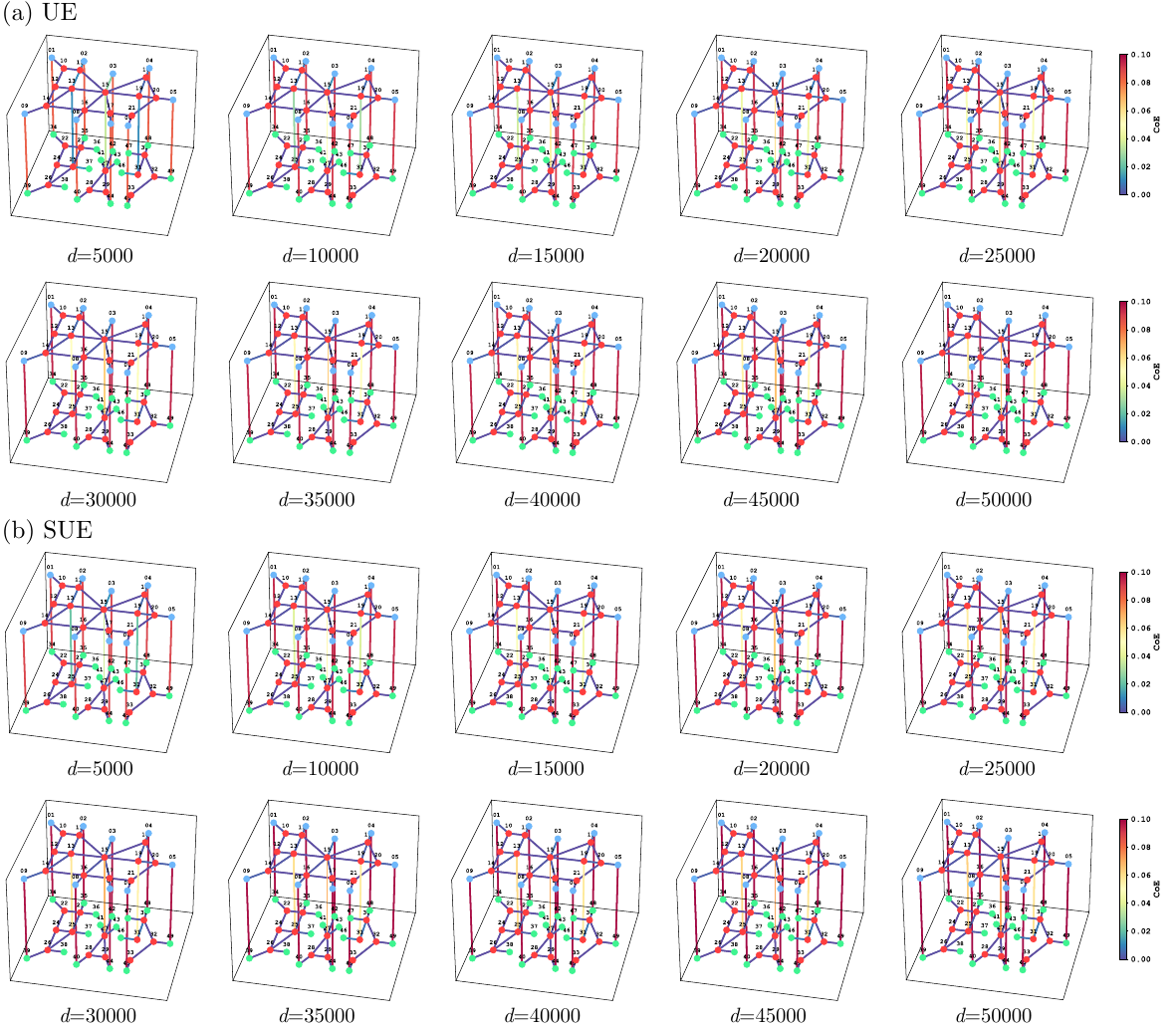}
\caption{CoE of evacuation network under varying evacuation demand: (a) UE configurations, (b) SUE configurations.}
\label{fig10}
\end{figure}

Fig.\ref{fig10} illustrates the variation of CoE across network edges as evacuation demand increases under both UE and SUE configurations. In the UE configuration, the edges that vertically connect the origin and destination nodes exhibit the highest CoE values, while the remaining edges display comparatively lower CoE levels. With rising demand, the CoE of $e_{13,25}$, $e_{15,27}$, and $e_{19,31}$ increases rapidly, whereas the horizontally oriented edges show only marginal growth. Under the SUE configuration, however, the CoE of edges throughout the network is systematically higher than under the UE configuration, with $e_{13,25}$, $e_{15,27}$, and $e_{19,31}$ already attaining relatively high CoE values even at low demand levels. In all configurations, the CoE values of the network edges remain greater than or equal to zero, indicating that, under the given parameters and conditions, neither the UE nor the SUE configuration supports the occurrence of Braess’ paradox.

\begin{figure}[ht!]
\centering
\includegraphics[scale=0.75]{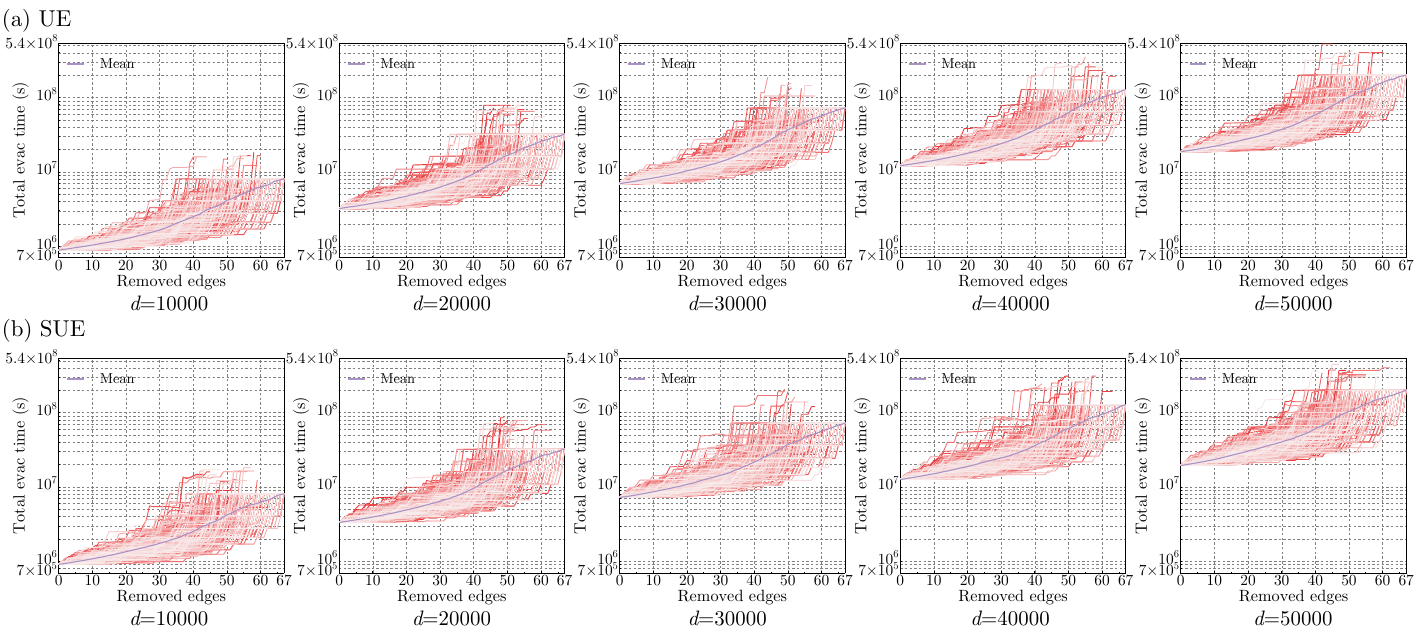}
\caption{Variation of total evacuation time under different demand levels with successive random edge removals: (a) UE configurations, (b) SUE configurations.}
\label{fig11}
\end{figure}

Although no occurrence of Braess' paradox was observed in the CoE evaluation based on isolated edges, the situation may differ under emergency conditions such as fire or terrorist attacks, where the topology of the evacuation network is systematically disrupted. In such circumstances, the possibility of Braess' paradox remains an important question. From this perspective, the potential for Braess' paradox in a damaged network topology was investigated by repeatedly removing edges at random. Following the same configuration as above, the demand was increased in increments of 10,000, with the maximum demand fixed at 50,000. Based on Monte Carlo simulations, edges were randomly removed, and the variation in total evacuation time was computed. A single simulation was terminated once the network was reduced to a state with no available evacuation paths. For each demand level, 500 Monte Carlo simulations were conducted, and the results are presented in Fig.\ref{fig11}.

Fig.\ref{fig11} illustrates the variation in the total system evacuation time under UE and SUE configurations when network edges are randomly removed. The purple curve represents the averaged results of 500 Monte Carlo simulations, which can be regarded as the statistical trend of system evolution. The mean line exhibits a stable upward trajectory, and since the figure adopts a semi-logarithmic scale, this indicates that the total evacuation time increases exponentially as more edges are removed. This finding reveals a phenomenon whereby, in a disrupted evacuation network, each additional removal of an edge statistically results in more severe consequences.  

By examining the variations of individual curves, the occurrence of Braess’ paradox can be identified by whether the total evacuation time does decrease when an edge is removed. In the UE configuration, this trend remains stable, as no cases were observed in which the removal of an edge led to a reduction in the total evacuation time. In contrast, under the SUE configuration, Braess’ paradox was observed: after ruling out the possibility of numerical errors, it was found that in certain instances, the removal of an edge from the disrupted network led to a reduction in the total evacuation time.

\section{Discussion and Conclusions} \label{section5}

In this study, the question of whether Braess' paradox exists in pedestrian evacuation traffic systems was examined through both experimental investigation and model-based analysis. Initially, the paradox appeared conceptually intriguing, and we hypothesized that such a phenomenon might frequently occur in practice. However, as the research progressed, this perspective was gradually reconsidered.

Based on the experimental findings, no evidence was obtained to support the occurrence of Braess' paradox. Given the constraints of the experimental setup and evacuation demand, this result was to be expected. Subsequently, the characteristics observed in the experiments were transferred into the corresponding UE and SUE models through parameter calibration, representing traffic assignment under perfect knowledge and limited knowledge, respectively. The simulation results demonstrated that, under conditions of high evacuation demand (typically corresponding to emergency scenarios), Braess' paradox is unlikely to occur when complete network information is assumed (UE assignment). By contrast, when stochastic errors in path information are incorporated, the paradox may emerge.

The distinction between UE and SO in the context of pedestrian evacuation illustrates the divergence between individual selfish decisions and centralized management. Nevertheless, under the specified parameters and assumptions, the model results indicated that attempts to approximate the system optimal flow pattern by blocking certain links were often counterproductive in practice. Since evacuation demand in pedestrian systems constitutes an essential precondition for the occurrence of Braess’ paradox, it is not fixed during emergencies. Therefore, expectations that the closure of specific corridors would improve overall evacuation efficiency are not warranted.

Several limitations of this study should be acknowledged. First, in the pedestrian evacuation experiments, only a single trial was conducted for each experimental condition. This design was primarily intended to avoid the potential influence of fatigue on evacuation speed; however, it may also amplify the impact of randomness on route choice. Second, because the experiment under S1 was conducted first, evacuees became more familiar with $p_1$ and $p_2$, which evidently influenced their route choice behavior in S2. This partly explains why the UE model exhibited higher accuracy than the SUE model in predicting the path ratio (see Fig.\ref{fig5}(b) in Section \ref{subsection3.1}). Finally, the inherent IIA issue of the SUE model is expected to result in limitations in flow assignment and is also anticipated to exert some impact on the accuracy of the results.

\centerline{}
\section*{Data Availability}
Code and data can be found at: \url{https://drive.google.com/drive/folders/17uO0Cyo2Q9659Ry4fLPSN-v4_JfG3feC} (Google Drive).

\centerline{}
\section*{Acknowledgments}
This work was supported by the National Natural Science Foundation of China (Grant No. 52072286, 71871189, 51604204), and the Fundamental Research Funds for the Central Universities (Grant No. 2022IVA108).

\appendix

\section{Experimental setup} \label{Appendix A}

The purpose of the experiment was to observe evacuees’ route-choice behavior under emergency egress conditions. The detailed experimental configuration is summarized in Table \ref{table2}. All participants were recruited from Wuhan University of Technology and were 21–26 years old. The mean age of male participants was 23.00 years and that of female participants was 22.53 years. The mean standing height measured with footwear was 175.10 cm for males and 163.13 cm for females. The mean body mass was 72.02 kg for males and 54.97 kg for females. Prior to data collection, all participants traversed the entire experimental area to familiarize themselves with the setting. During the trials, participants were instructed to evacuate to the designated destination area as quickly as possible upon the start signal. Video for observation was captured by a camera mounted on the roof of an adjacent building at a height of approximately 25 m, with a resolution of 1920 × 1080 pixels at 25 frames per second. Trajectories were extracted semi-automatically using the PeTrack software. 

\begin{figure}[ht!]
\centering
\includegraphics[scale=0.5]{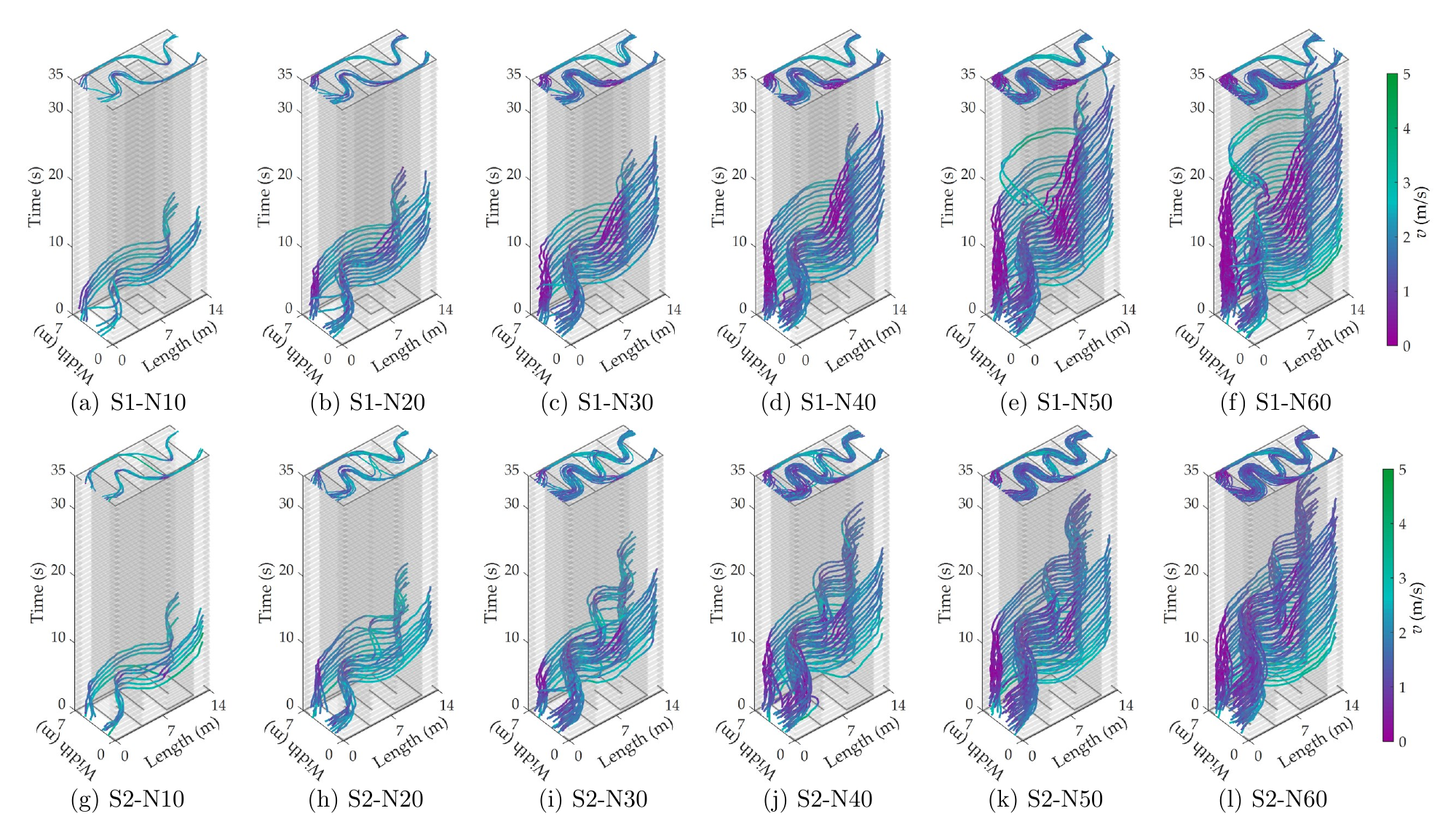}
\caption{Spatiotemporal evacuation trajectories: the upper part presents data from S1, while the lower part presents data from S2.}
\label{fig12}
\end{figure}

\begin{table}[htbp]
\centering
\caption{Experiment settings}
\label{table2} 
\begin{tabular}{ccccc}
\toprule
\textbf{Index} & \textbf{Scenario} & \textbf{Participants} & \textbf{Ratios (male/female)} & \textbf{Repetitions} \\
\midrule
S1-N10  & 1 & 10  & 5/5   & 1 \\
S1-N20  & 1 & 20  & 10/10 & 1 \\
S1-N30  & 1 & 30  & 15/15 & 1 \\
S1-N40  & 1 & 40  & 20/20 & 1 \\
S1-N50  & 1 & 50  & 25/25 & 1 \\
S1-N60  & 1 & 60  & 30/30 & 1 \\
S2-N10  & 2 & 10  & 5/5   & 1 \\
S2-N20  & 2 & 20  & 10/10 & 1 \\
S2-N30  & 2 & 30  & 15/15 & 1 \\
S2-N40  & 2 & 40  & 20/20 & 1 \\
S2-N50  & 2 & 50  & 25/25 & 1 \\
S2-N60  & 2 & 60  & 30/30 & 1 \\
\bottomrule
\end{tabular}
\end{table}

Fig.\ref{fig12} presents the spatiotemporal trajectories of evacuees within the experimental area for the groups corresponding to S1 and S2. The geometric configuration of the experimental region is clearly illustrated, corresponding to the 2-path and 4-path networks. In addition, the distribution of evacuation speeds and the proportions of path choices are depicted.

\begin{figure}[ht!]
\centering
\includegraphics[scale=0.8]{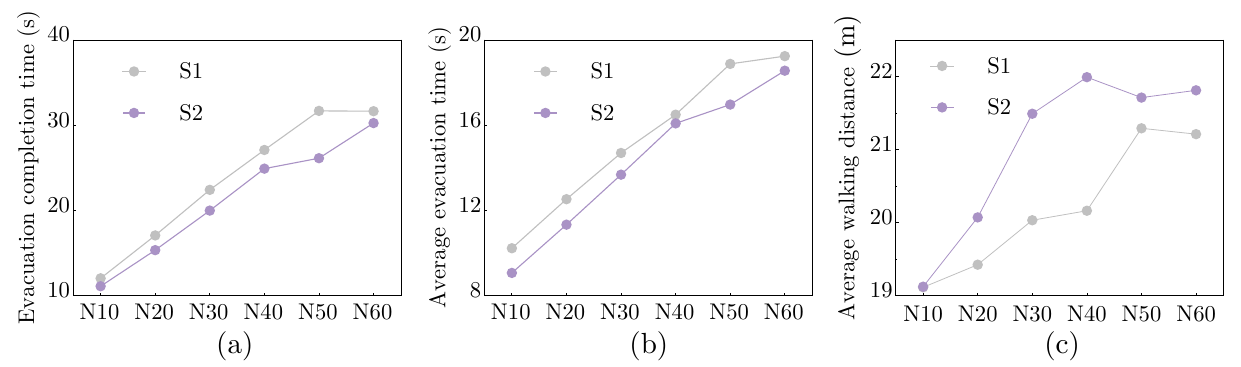}
\caption{Trends of basic experimental data with varying evacuation demand: (a) evacuation completion time, (b) average evacuation time, (c) average walking distance.}
\label{fig13}
\end{figure}

Fig.\ref{fig13} presents the variations in evacuation completion time, mean evacuation time, and mean walking distance with increasing evacuation demand under scenarios S1 and S2. Here, the evacuation completion time is defined as the time interval between the entry of the first evacuee into the experimental area and the exit of the last evacuee. It can be observed that, in the 4-path network, both the evacuation completion time and the mean evacuation time are shorter than those in the 2-path network. These data present negative empirical findings regarding Braess’ paradox in pedestrian evacuation traffic.

\begin{figure}[ht!]
\centering
\includegraphics[scale=0.42]{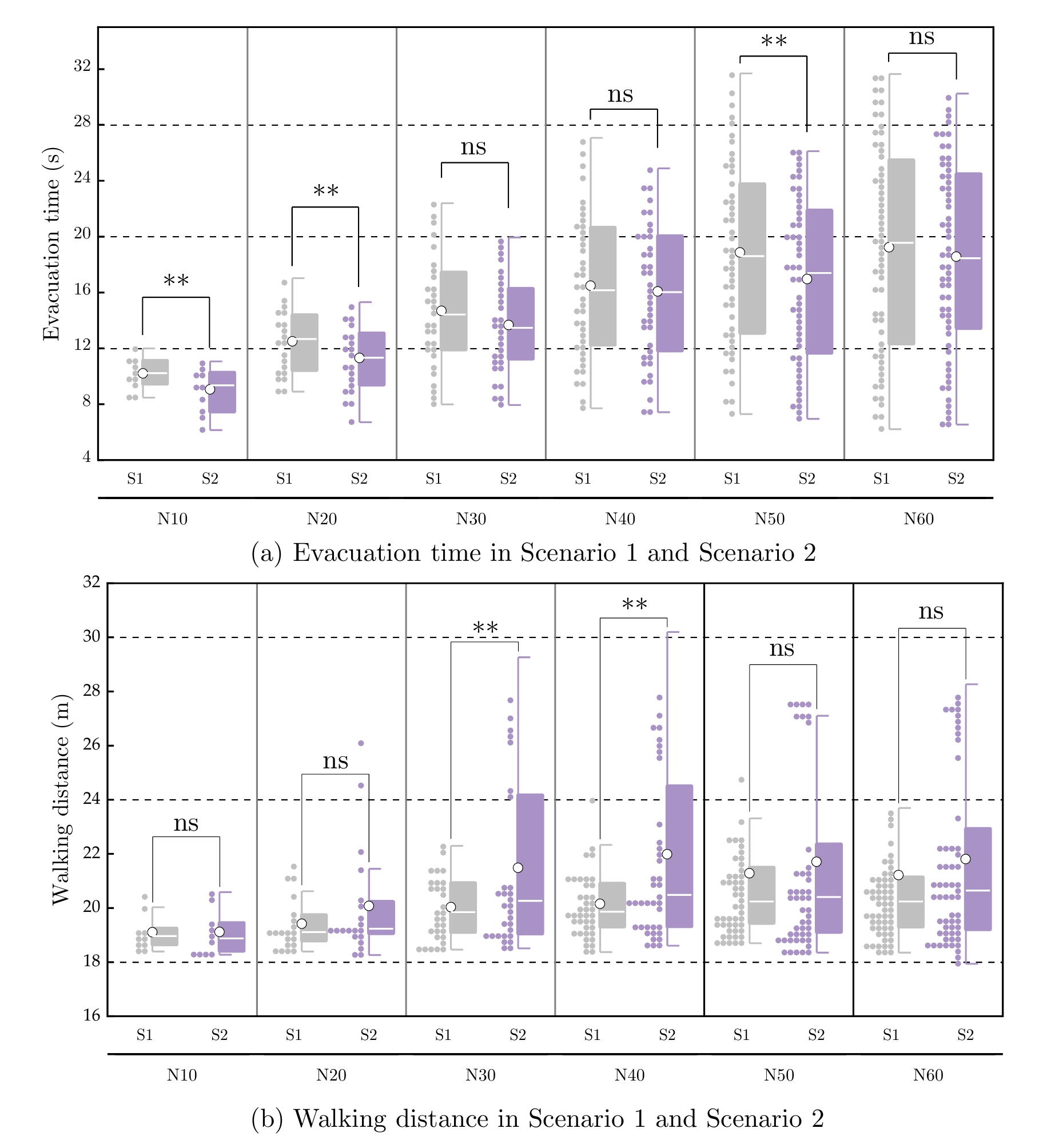}
\caption{Distributions of evacuation time (a) and walking distance (b) under varying evacuation demands, with non-parametric test results (Wilcoxon test, **: \(p\)<0.01) indicated by the lines.}
\label{fig14}
\end{figure}

Fig.\ref{fig14} presents the distributions and comparisons of evacuation time and walking distance per evacuee under different evacuation demand levels in the S1 and S2 experimental scenarios. The dashed lines denote the results of significance tests for the differences between the S1 and S2 groups. It can be observed that, in many sets, the evacuation data did not exhibit statistically significant differences at the 0.01 level.

\bibliographystyle{aasjournal}

\end{document}